\documentclass[twocolumn]{jpsj2}
\def\runtitle{On Superconducting Double Transition in ${\rm PrOs_4Sb_{12}}$}

\def\runauthor{
Masanori {\sc Ichioka}, Noriyuki {\sc Nakai} and Kazushige {\sc Machida}
}

\title{%
On Superconducting Double Transition in ${\rm PrOs_4Sb_{12}}$}

\author{%
Masanori {\sc Ichioka},\footnote{E-mail address: oka@mp.okayama-u.ac.jp} 
Noriyuki {\sc Nakai} and Kazushige {\sc Machida}
}

\inst{%
Department of Physics, Okayama University, Okayama 700-8530}

\recdate{
March 20, 2003
}

\abst{%
Superconducting double transition in ${\rm PrOs_4Sb_{12}}$ is 
investigated by analyzing the anisotropy of the 
upper critical field $H_{\rm c2}$ in the $ab$-plane, 
and the possible pairing state is discussed. 
When mixing due to gradient coupling is active, 
the twofold-symmetric component is necessarily induced in the fourfold 
symmetric phase, leading to the twofold oscillation of $H_{\rm c2}$, 
contrary to the experimental result. 
To avoid the mixing effect, the weak spin-orbit coupling 
triplet pairing state is considered as a likely pairing function, 
where time-reversal symmetry is broken.  
} 

\kword{%
$H_{c2}$ anisotropy, 
unconventional superconductor, 
pairing function, 
${\rm PrOs_4Sb_{12}}$
}

\begin{document}
%

\sloppy
\maketitle




The newly discovered superconductor ${\rm PrOs_4Sb_{12}}$ 
is a heavy-fermion compound with filled skutterudite 
structure.~\cite{Bauer} 
It is considered that quadrupole fluctuation is related to 
the pairing of superconductivity, suggesting a new pairing mechanism.  
In order to study the pairing mechanism, 
it is important to identify the pairing symmetry. 
The experiment on the nuclear spin-lattice relaxation rate 
suggested that ${\rm PrOs_4Sb_{12}}$ has full gap or point nodes, 
excluding line nodes, at a zero field.~\cite{Kotegawa} 
The unique phenomenon in the study of the pairing symmetry 
is the superconducting double transition. 
The specific heat jump at the transition temperature $T_{\rm c}$ 
is broad, suggesting a double transition.~\cite{Vollmer} 
By the thermal transport measurement in the magnetic field ($H$) rotated 
in the $ab$-plane of the crystal axes, the phase diagram of the 
double transition was revealed.\cite{Izawa}  
This experiment also suggested that a fourfold-symmetric pairing function 
around the $c$-axis in the high-field H-phase is changed 
to a twofold-symmetric one in the low-field L-phase at the second transition, 
when magnetic field or temperature is decreased. 
Thus, the $H$-$T$ phase diagram is divided into two regions,  
H- and L-phases. 

To explain the double transition, we need multiple components  
for the pairing functions, i.e., an unconventional pairing 
other than the conventional $s$-wave pairing.  
The fourfold-twofold transition in ${\rm PrOs_4Sb_{12}}$ 
can be explained as follows: 
After the fourfold-symmetric pairing component appears at the first transition,
the second component appears at the second transition, 
and the combination of these components gives the twofold-symmetric 
gap structure. 
For ${\rm PrOs_4Sb_{12}}$, 
the scenario of ``anisotropic-$s$''+${\rm i}d$-wave pairing~\cite{Goryo}  
with point nodes\cite{Maki} was proposed.
This can explain the double transition at zero field under the cubic symmetry. 
However, in this scenario, it is difficult to reproduce the 
fourfold-twofold transition in magnetic fields, as will be discussed later. 

In the vortex state at a finite magnetic field, 
we have to consider the mixing of the multiple components of the 
pairing function. 
In the multicomponent Ginzburg-Landau (GL) theory, 
this mixing comes from the gradient coupling between the 
different pairing components. 
This mixing effect has been well studied in the $d$-wave 
pairing case for high-$T_{\rm c}$ superconductors,
\cite{Sugiyama,IchiokaEnomoto,IchiokaDmix,IchiokaDlattice,Ren1,Berlinsky,FranzKallin}  
and in the triplet pairing case for ${\rm Sr_2RuO_4}$ and 
${\rm UPt_3}$.\cite{Agterberg,TakigawaP,IchiokaP,MachidaUPt3,MachidaUPt3v,Fujita}
In the magnetic field, if the mixing effect is active, 
a twofold component is induced even above the fourfold-twofold 
transition, indicating that the resulting superconducting gap shows a 
twofold-symmetric character and that the phase transition 
between the H- and L-phases is smeared. 
The easiest way to clarify the effect of the twofold-symmetric character 
is to study $H_{\rm c2}$ anisotropy in the magnetic field rotated in 
the $ab$-plane. 
If $H_{\rm c2}$ has twofold anisotropy, all quantities should 
show the twofold anisotropy near $H_{\rm c2}$. 
Since twofold anisotropy is not observed in the experiment, 
we have to consider the condition that it does not occur. 
This gives important information for identifying the pairing function for 
${\rm PrOs_4Sb_{12}}$. 
Also in ${\rm UPt_3}$, a theoretical study to explain the phase diagram 
of the double transition in the magnetic field strongly suggested 
triplet pairing.\cite{MachidaUPt3}  

The purpose of this letter is to study $H_{\rm c2}$ anisotropy for the 
fourfold-twofold transition case, and to estimate the effect of the 
mixing between fourfold and twofold pairing components. 
By obtaining an analytical result for $H_{\rm c2}$ near $T_{\rm c}$ 
in the microscopic theory, 
we show generally that $H_{\rm c2}$ shows twofold oscillation, 
contrary to the experimental result,\cite{Izawa} 
if gradient coupling terms are included. 
Therefore, gradient coupling terms should be absent 
in order to assure the fourfold behavior above the second transition.  
Our calculation is performed using the general pairing function forms 
for both singlet and triplet pairings, 
so that our analysis can be applied to any pairing function cases. 
In the last part, we discuss the possible pairing function for 
${\rm PrOs_4Sb_{12}}$. 

For the multicomponent superconductors, the pair potential is decomposed to 
\begin{eqnarray} 
\Delta(\mib{r},\mib{ k})_{\alpha\beta} 
=\sum_m \eta_m(\mib{ r}) \hat{\phi}_m(\mib{ k})_{\alpha \beta }
\end{eqnarray} 
with the order parameter $\eta_m(\mib{ r})$ for the $m$-th pairing 
component.\cite{Sigrist}  
Here, $\mib{ r}$ is the center of mass coordinate of the Cooper pair. 
The relative momentum $\mib{ k}$ of the pair is mapped on the 
Fermi surface. 
The gap function is a $2 \times 2$ matrix of $\alpha$ and $\beta$ 
($=\uparrow$, $\downarrow$), given by 
\begin{eqnarray}  
\hat\phi_m(\mib{ k})_{\alpha\beta}
=
\left( \begin{array}{cc} 
0 &  \phi_m(\mib{ k})\\ 
-\phi_m(\mib{ k}) & 0 \\
\end{array} \right)
\end{eqnarray} 
for singlet pairing, and 
\begin{eqnarray}   &&  \hspace{-0.8cm}
\hat\phi_m(\mib{ k})_{\alpha\beta}
\nonumber \\ && \hspace{-0.8cm}
=
\left( \begin{array}{cc} 
-d_{m,x}(\mib{ k}) +{\rm i}d_{m,y}(\mib{ k}) & d_{m,z}(\mib{ k}) \\ 
d_{m,z}(\mib{ k})  & d_{m,x}(\mib{ k}) +{\rm i}d_{m,y}(\mib{ k}) \\
\end{array} \right)
\qquad 
\end{eqnarray} 
for triplet pairing. 
The pairing function satisfies the normalization and orthogonalization 
condition $ \langle \overline{\phi_m^\ast \phi_{m'} } \rangle_{\mib{k}}
=\delta_{m,m'}$, where 
$\overline{\phi_m^\ast \phi_{m'} }
\equiv 
{\rm tr}(\hat{\phi}_m^\dagger \hat{\phi}_{m'})/2
=\phi_m^\ast \phi_{m'} $ 
for singlet pairing and 
$\overline{\phi_m^\ast \phi_{m'} }=
d_{m,x}^\ast d_{m',x} +d_{m,y}^\ast d_{m',y} +d_{m,z}^\ast d_{m',z} 
=\mib{ d}_m^\ast \cdot \mib{ d}_{m'}$ 
for triplet pairing. 
$\langle \cdots \rangle_{\mib{k}}$ means the average of $\mib{ k}$ 
on the Fermi surface. 
Similarly, the pairing interaction is decomposed to 
\begin{eqnarray} 
V(\mib{ k}',\mib{ k})_{\alpha'\beta',\alpha\beta}
=\sum_m V_m \hat{\phi}_m(\mib{ k}')_{\alpha'\beta'}
            \hat{\phi}_m^\dagger(\mib{ k})_{\alpha \beta }. 
\end{eqnarray} 
The transition temperature $T_{{\rm c},m}$ for the $m$-th component 
is given by the relation $(V_m N_{\rm F})^{-1}
=\ln(2 \omega_{\rm c} \bar{\gamma}/\pi T_{{\rm c},m})$ 
with the Euler constant $\bar{\gamma}$ 
and the cutoff frequency $\omega_{\rm c}$. 
$N_{\rm F}$ is the averaged density of states on the Fermi surface. 

We derive the gap equation along the $H_{\rm c2}$ line 
by the quasiclassical method.\cite{Schopohl} 
The $2 \times 2$ matrix quasiclassical Green's functions $f$ and $g$ 
are determined using the quasiclassical Eilenberger equation 
\begin{eqnarray} &&
(2\omega_n +{\rm i}\mib{ v}\cdot\mib{ q})f
=(g \Delta+\Delta\bar{g}), 
\nonumber \\ && 
(2\omega_n +{\rm i}\mib{ v}\cdot\mib{ q}^\ast)\bar{f}
=(\bar{g} \Delta^\dagger +\Delta^\dagger g ), 
\label{eq:Eilenberger}
\end{eqnarray} 
where $g^2 + f \bar{f}=\hat{1}$ and  
$\bar{g}^2 + \bar{f}f =\hat{1}$ with 
a unit matrix $\hat{1}$. 
In eq. (\ref{eq:Eilenberger}), $\mib{ v}$ is the Fermi velocity, and 
$\mib{ q}=-{\rm i}(\partial/\partial \mib{ r}) 
       +(2 \pi/\phi_0)\mib{ A}(\mib{ r}) $
with a vector potential $\mib{ A}(\mib{ r})=(-Hy,0,0)$ 
in Landau gauge and the flux quantum $\phi_0$. 
The selfconsistent condition is given by 
\begin{eqnarray} &&
\Delta(\mib{ r},\mib{ k}')_{\alpha'\beta'}
=\frac{1}{2}\sum_{\alpha\beta} \pi T \sum_{|\omega_n|<\omega_{\rm c}} \langle  
N_{\rm F} V(\mib{ k}',\mib{ k})_{\alpha'\beta',\alpha\beta}
\nonumber \\ && \hspace{4cm} \times 
f(\mib{ r},\mib{ k},{\rm i}\omega_n)_{\alpha\beta} \rangle_{\mib{k}} . 
\label{eq:gapeq1}
\end{eqnarray} 
After eq. (\ref{eq:Eilenberger}) is linearized about $\Delta$ along 
$H_{c2}$, we expand $f$ in powers of $\mib{ v}\cdot\mib{ q}$,\cite{Schopohl}  
giving 
$f_{\alpha \beta }=
\sum_{j=0}^\infty \{ -{\rm i}{\rm sgn}(\omega_n)/2 \}^j 
|\omega_n|^{-(j+1)} (\mib{ v}\cdot\mib{ q})^j 
\Delta_{\alpha \beta } $.
By substituting this in eq. ({\ref{eq:gapeq1}), we obtain 
the gap equation 
\begin{eqnarray} && \hspace{-0.8cm}
\eta_{m'} = N_{\rm F} V_{m'} \sum_{n=0}^{\infty} A_{2n} \sum_m 
\langle (\mib{ v}\cdot\mib{ q})^{2n}  
\overline{\phi_{m'}^\ast \phi_m } \rangle_{\mib{k}}  
\eta_m  , 
\label{eq:gapeq2}
\end{eqnarray} 
with 
$A_{2n}= 2(-1)^n (2 \pi T)^{-2n}( 1- 2^{-(2n+1)} ) \zeta(2n+1) $ 
for $n>0$, and 
$A_0=\ln(2 \omega_{\rm c}\bar{\gamma}/\pi T) 
=(V_1 N_{\rm F})^{-1}+ \ln(T_{\rm c,1}/T)$. 

We assume that the first component $\eta_1$ is the dominant one with  
the highest transition temperature $T_{\rm c}=T_{{\rm c},1}$, 
and that other components $(m \ge 2)$ appear at the lower temperature 
$T_{{\rm c},m}< T_{{\rm c}}$, 
i.e., double transition occurs at zero field. 
We solve eq. (\ref{eq:gapeq2}) in perturbation of $\ln(T_{\rm c}/T)$, 
and obtain $H_{\rm c2}$ near $T_{{\rm c}}$.\cite{Sugiyama,IchiokaEnomoto} 
By introducing annihilation and creation operators for the Landau level as 
$a =-\epsilon^{-1/2}(q_x - {\rm i} \nu^{-1}q_y)$ and 
$a^\dagger=-\epsilon^{-1/2}(q_x + {\rm i} \nu^{-1}q_y)$, respectively,  
we can write 
$ \mib{ v}\cdot\mib{ q}
=-v_{\rm F}\sqrt{\epsilon}(\hat{v}_+ a + \hat{v}_- a^\dagger)$, 
where $\epsilon=4 \pi H/\phi_0\nu$,  
$\hat{v}_{\pm} =(\hat{v}_x \pm {\rm i}\nu \hat{v}_y )/2$, 
$\hat{\mib{v}}=\mib{ v}/v_{\rm F}$ 
with the averaged Fermi velocity $v_{\rm F}$. 
Since we assume that $z$-axis is parallel to the applied field, 
$q_z=0$. 
The anisotropy ratio $\nu$ is defined as 
$\nu = (\langle \hat{v}_x^2 
         \overline{|\phi_1|^2} \rangle_{\mib{k}}/ 
        \langle \hat{v}_y^2 
         \overline{|\phi_1|^2} \rangle_{\mib{k}})^{1/2}$ 
with $\overline{|\phi_1|^2}\equiv \overline{\phi_1^\ast \phi_1}$, 
so that 
$ \langle \hat{v}_\pm^2 \overline{|\phi_1|^2} \rangle_{\mib{k}}=0$. 
In dimensionless form, eq. (\ref{eq:gapeq2}) is written as 
\begin{eqnarray} && \hspace{-0.8cm}
0=\eta_1 
- 2 \xi^2\epsilon \sum_m 
 \langle (\hat{v}_+ a + \hat{v}_- a^\dagger)^{2} 
     \overline{\phi_1^\ast \phi_m} \rangle \eta_m 
\nonumber \\ &&  \hspace{-0.8cm} 
+2\gamma (\xi^2 \epsilon)^2  \sum_m 
 \langle (\hat{v}_+ a + \hat{v}_- a^\dagger)^{4} 
     \overline{\phi_1^\ast \phi_m} \rangle \eta_m
+\cdots, 
\label{eq:gapeq3a}
\\ &&  \hspace{-0.8cm}
0= - \eta_{m'}
- 2 \gamma c_{m'} \xi^2 \epsilon \sum_m 
\langle (\hat{v}_+ a + \hat{v}_- a^\dagger)^{2} 
     \overline{\phi_{m'}^\ast \phi_m} \rangle \eta_m
\nonumber \\ &&  \hspace{-0.8cm} 
+ 2 \gamma^2 c_{m'} (\xi^2 \epsilon)^2 \sum_m 
\langle (\hat{v}_+ a + \hat{v}_- a^\dagger)^{4} 
     \overline{\phi_{m'}^\ast \phi_m} \rangle \eta_m
+\cdots 
\qquad 
\label{eq:gapeq3b}
\end{eqnarray}
for $m' \ge 2$, 
where 
$\gamma=\alpha \ln(T_{\rm c}/T)$,  
$\alpha=2 |A_4|/|A_2|^2=0.908$, 
$\xi=(|A_2| v_{\rm F}^2 / 2 \ln(T_{\rm c}/T))^{1/2}$ and 
$1/c_m=\alpha \ln(T/T_{{\rm c},m})$. 
Equations (\ref{eq:gapeq3a}) and (\ref{eq:gapeq3b}) 
include the gradient coupling terms between 
different components and the nonlocal correction terms in the GL theory. 
In terms of the Landau level function 
$ \psi_n(\mib{ r})=(n!)^{-1/2}(a^\dagger)^n \psi_0(\mib{ r})$ 
for the $n$-th Landau level, the order parameters are decomposed to 
$\eta_m(\mib{ r})=\sum_{n=0}^{\infty}\eta_{m,n}\psi_n(\mib{ r})$. 
By expanding  $\eta_{m,n}$ and $\epsilon$ in powers of $\gamma$, 
we can solve eqs. (\ref{eq:gapeq3a}) and (\ref{eq:gapeq3b}) in perturbation. 
As a result, we obtain 
\begin{eqnarray} && 
H_{c2}=\frac{\phi_0 }{4 \pi \xi^2} \left(
\langle \hat{v}_x^2 \overline{|\phi_1|^2} \rangle_{\mib{k}} 
\langle \hat{v}_y^2 \overline{|\phi_1|^2} \rangle_{\mib{k}} \right)^{-1/2}
\nonumber \\ && \times 
\{  
1+ \gamma ( \tilde{\epsilon}_1 
+\sum_{m \ge 2}  c_m \tilde{\epsilon}_m ) +O(\gamma^2) \} 
\label{eq:hc2a}
\end{eqnarray}
with 
\begin{eqnarray}  && \hspace{-0.8cm}
\tilde{\epsilon}_1
=\frac
{3\langle \hat{v}_+^2 \hat{v}_-^2 \overline{|\phi_1|^2} \rangle_{\mib{k}}  }
{2\langle \hat{v}_+ \hat{v}_- \overline{|\phi_1|^2} \rangle_{\mib{k}}^2 } , 
\label{eq:hc2b}
\\ && \hspace{-0.8cm}
\tilde{\epsilon}_m
=\frac{ |\langle \hat{v}_+ \hat{v}_- 
     \overline{\phi_1^\ast \phi_m} \rangle_{\mib{k}} |^2 
      +2|\langle \hat{v}_+^2 
     \overline{\phi_1^\ast \phi_m} \rangle_{\mib{k}} |^2 }
      {\langle \hat{v}_+ \hat{v}_- 
     \overline{|\phi_1|^2} \rangle_{\mib{k}} ^2 } 
. \quad (m \ge 2) \qquad 
\label{eq:hc2c} 
\end{eqnarray} 
Along $H_{\rm c2}$, the order parameters are given by 
\begin{eqnarray}&&  \hspace{-0.8cm}
\eta_1(\mib{{r}}) =\eta_{1,0}[ \psi_0(\mib{{r}}) 
+\gamma \{ \tilde{\eta}_{1,2}\psi_2(\mib{{r}}) 
          +\tilde{\eta}_{1,4}\psi_4(\mib{{r}})\} +O(\gamma^2) ]  , 
\nonumber\\ && \qquad 
\\ && \hspace{-0.8cm}
\eta_m(\mib{{r}}) =c_m \eta_{1,0}[ 
 \gamma \{ \tilde{\eta}_{m,0}\psi_0(\mib{{r}}) 
          +\tilde{\eta}_{m,2}\psi_2(\mib{{r}}) \} +O(\gamma^2)]  
\qquad 
\label{eq:opm}
\end{eqnarray}
for $m \ge 2$. 
The factors $\tilde{\eta}_{m,n}$ are obtained as in 
eqs. (2.12)-(2.16) of ref. \citen{Sugiyama}. 

We consider the case when a magnetic field is applied parallel to 
the $ab$-plane of the crystal coordinate. 
In this case, the transformation of the Fermi velocity 
from $(\hat{v}_a,\hat{v}_b,\hat{v}_c)$ in the crystal coordinate 
to $(\hat{v}_x,\hat{v}_y,\hat{v}_z)$ in the vortex coordinate 
is given as 
$ (\hat{v}_x, \hat{v}_y, \hat{v}_z) = 
(-\hat{v}_a \sin\theta +\hat{v}_b \cos\theta , \hat{v}_c , 
 \hat{v}_a \cos\theta +\hat{v}_b \sin\theta )$, 
where $\theta$ is the rotating angle of the magnetic field in the $ab$-plane. 
We assume that the superconducting gap 
$\overline{|\phi_1|^2}$ for the dominant component 
has a fourfold-symmetry around the $c$-axis, so that 
$\langle \hat{v}_a^2 \overline{|\phi_1|^2} \rangle_{\mib{k}}
=\langle \hat{v}_b^2 \overline{|\phi_1|^2} \rangle_{\mib{k}}$ and 
$\langle \hat{v}_a \hat{v}_b \overline{|\phi_1|^2} \rangle_{\mib{k}} =0 $. 
Therefore, we obtain 
$\langle \hat{v}_x^2 \overline{|\phi_1|^2} \rangle_{\mib{k}}
=\langle \hat{v}_a^2 \overline{|\phi_1|^2} \rangle_{\mib{k}}$,  
$\langle \hat{v}_y^2 \overline{|\phi_1|^2} \rangle_{\mib{k}}
=\langle \hat{v}_c^2 \overline{|\phi_1|^2} \rangle_{\mib{k}}$, 
$\nu^2=\langle \hat{v}_a^2 \overline{|\phi_{1}|^2} \rangle_{\mib{k}} 
/\langle \hat{v}_c^2 \overline{|\phi_{1}|^2} \rangle_{\mib{k}} $ 
and 
$\langle \hat{v}_+ \hat{v}_- \overline{|\phi_1|^2} \rangle_{\mib{k}}^2   
=\langle \hat{v}_a^2 \overline{|\phi_1|^2} \rangle_{\mib{k}}^2/4$.  
As a result, 
\begin{eqnarray} && \hspace{-0.8cm}
\tilde{\epsilon}_1 
=\frac{3}{32} 
\langle \hat{v}_a^2 \overline{|\phi_1|^2} \rangle_{\mib{k}}^{-2} 
\Bigl\{ 
4 \langle(\hat{v}_a^2 +\nu^2 \hat{v}_c^2)^2 
              \overline{|\phi_1|^2} \rangle_{\mib{k}}  
\nonumber \\ && \hspace{-0.8cm}\hspace{0.5cm} 
  -\langle(\hat{v}_a^4 -3 \hat{v}_a^2 \hat{v}_c^2)
    \overline{|\phi_1|^2} \rangle_{\mib{k}}  
 ( 1-\cos 4\theta )
\Bigr\} , 
\qquad 
\label{eq:hc2e1}
\\ && \hspace{-0.8cm}
\tilde{\epsilon}_m 
= \frac{1}{16} 
\langle \hat{v}_a^2 \overline{|\phi_1|^2} \rangle_{\mib{k}}^{-2}  
\Bigl\{  
\Bigl|
  \langle (\hat{v}_a^2 +\hat{v}_b^2 +2 \nu^2 \hat{v}_c^2 )
  \overline{\phi_{1}^\ast \phi_m}  \rangle_{\mib{k}}
\nonumber \\ && \hspace{-0.8cm}\hspace{0.5cm} 
- \langle (\hat{v}_a^2 -\hat{v}_b^2)
  \overline{\phi_{1}^\ast \phi_m}  \rangle_{\mib{k}} \cos 2\theta 
 -2\langle \hat{v}_a \hat{v}_b 
  \overline{\phi_{1}^\ast \phi_m} \rangle_{\mib{k}}  \sin 2\theta
\Bigr|^2 
\nonumber \\ && \hspace{-0.8cm}\hspace{0.5cm} 
+2\Bigl|
  \langle (\hat{v}_a^2 +\hat{v}_b^2 -2 \nu^2 \hat{v}_c^2 )
  \overline{\phi_{1}^\ast \phi_m}  \rangle_{\mib{k}}
\nonumber \\ && \hspace{-0.8cm}\hspace{0.5cm} 
- \langle (\hat{v}_a^2 -\hat{v}_b^2)
  \overline{\phi_{1}^\ast \phi_m}  \rangle_{\mib{k}} \cos 2\theta 
 -2\langle \hat{v}_a \hat{v}_b 
  \overline{\phi_{1}^\ast \phi_m} \rangle_{\mib{k}}  \sin 2\theta
\nonumber \\ && \hspace{-0.8cm}\hspace{0.5cm} 
 \pm{\rm i}4 \nu 
\Bigl( \langle \hat{v}_b \hat{v}_c 
    \overline{\phi_{1}^\ast \phi_m} \rangle_{\mib{k}} 
                               \cos\theta 
 -\langle \hat{v}_a \hat{v}_c 
    \overline{\phi_{1}^\ast \phi_m} \rangle_{\mib{k}} 
                               \sin\theta  \Bigr)
\Bigr|^2 
\Bigr\} , \qquad 
\label{eq:hc2e2}
\end{eqnarray}
where  
$\langle  \hat{v}_a^4 \overline{|\phi_1|^2} \rangle_{\mib{k}} 
=\langle  \hat{v}_b^4 \overline{|\phi_1|^2} \rangle_{\mib{k}}$, 
$\langle  \hat{v}_a^2 \hat{v}_c^2 \overline{|\phi_1|^2} \rangle_{\mib{k}} 
=\langle  \hat{v}_b^2 \hat{v}_c^2 \overline{|\phi_1|^2} \rangle_{\mib{k}}$ 
and 
$ \langle  \hat{v}_a^3 \hat{v}_b \overline{|\phi_1|^2} \rangle_{\mib{k}}
= \langle  \hat{v}_a \hat{v}_b^3 \overline{|\phi_1|^2} \rangle_{\mib{k}} 
= \langle  \hat{v}_a \hat{v}_b \hat{v}_c^2 
\overline{|\phi_1|^2} \rangle_{\mib{k}} =0 $. 

From eqs. (\ref{eq:hc2a}), (\ref{eq:hc2e1}) and (\ref{eq:hc2e2}), 
we consider the $\theta$ dependence of 
$H_{\rm c2}$ when the magnetic field is rotated within the $ab$-plane. 
The term $\tilde{\epsilon}_1$  comes from the 
nonlocal correction term in the GL theory, and gives the fourfold 
anisotropy  of $H_{\rm c2}$ under field rotation because of $\cos 4\theta$ 
in eq. (\ref{eq:hc2e1}), reflecting fourfold-symmetry of the dominant pairing  
function $\hat\phi_1(\mib{{k}})$. 
This fourfold $H_{\rm c2}$ anisotropy was studied in the $d$-wave pairing 
case in high-$T_{\rm c}$ superconductors and the anisotropic $s$-wave pairing 
case in borocarbide superconductors.\cite{Sugiyama,Takanaka,Metlushko} 
The term $\tilde{\epsilon}_m$  comes from the 
mixing of $\hat\phi_1$ and $\hat\phi_m$ due to gradient coupling. 
By this mixing, $\eta_m(\mib{{r}})$ $(m \ge 2)$ is induced even 
at $T > T_{{\rm c},m}$ in the vortex state at a finite field, 
as shown in eq. (\ref{eq:opm}), while it does not appear at a zero field. 
When the pairing function $\hat{\phi}_m$ ($m \ge 2$) has twofold anisotropy 
to introduce the fourfold-twofold transition at a zero field,  
$\tilde{\epsilon}_m$ gives twofold anisotropy in the $\theta$ dependence 
of $H_{{\rm c2}}$. 
As an example, 
we consider the case of the ``anisotropic $s$''$+{\rm i}d$-wave singlet  
pairing scenario proposed for ${\rm PrOs_4Sb_{12}}$,\cite{Goryo}   
whose pairing functions are given as 
\begin{eqnarray} && 
\phi_1(\mib{{k}})=\frac{\sqrt{21}}{2}
                 \{ 1-(\hat{k}_a^4+\hat{k}_b^4+\hat{k}_c^4) \} , 
\label{eq:sg1}
\\ && 
\phi_2(\mib{{k}})=\frac{\sqrt{105}}{8}
                 ( 2 \hat{k}_b^4 -\hat{k}_c^4 -\hat{k}_a^4 ) , 
\\ && 
\phi_3(\mib{{k}})=\frac{3\sqrt{35}}{8}
                 (\hat{k}_c^4 -\hat{k}_a^4 ) 
\qquad 
\end{eqnarray} 
with $\hat{\mib{k}}=\mib{{k}}/k_{\rm F}$.\cite{Maki}    
Numerical factors come from the normalization condition  
$\langle |\phi_m|^2 \rangle_{\mib{{k}}}=1$. 
We assume an isotropic sphere Fermi surface. 
If $\eta_3$ or $\eta_2$ appears at the second transition into the L-phase, 
the fourfold-twofold transition is explained at a zero field 
within point-node pairing functions.  
However, we obtain 
\begin{eqnarray} && \hspace{-0.8cm}
\tilde{\epsilon}_1=\frac{1197}{715}+\frac{45}{572}(1-\cos4\theta), 
\\ && \hspace{-0.8cm}
\tilde{\epsilon}_2=\frac{169}{38720}
   \{ (1-3\cos2\theta)^2 +18(1+\cos2\theta)^2 \} , 
\\ && \hspace{-0.8cm}  
\tilde{\epsilon}_3=\frac{507}{38720}
   \{ (1+\cos2\theta)^2 +2 (3-\cos2\theta)^2 \}, 
\qquad 
\end{eqnarray} 
since $\langle (\hat{v}_a^2 -\hat{v}_b^2) 
                            \phi_1 \phi_m \rangle_{\mib{ k}}\ne 0$. 
Because of $\cos2\theta$ in $\tilde{\epsilon}_2$ and $\tilde{\epsilon}_3$, 
$H_{c2}$ shows twofold anisotropy, indicating that 
the H-phase becomes twofold-symmetric in the vortex state 
by the induced $\eta_2$ and $\eta_3$ components. 
This is contrary to the experimental result.\cite{Izawa}  
Approaching $T_{{\rm c},m}$, the twofold-symmetric contribution 
of $\tilde{\epsilon}_m$ increases because of the factor 
$c_m$ in eq. (\ref{eq:hc2a}). 

To suppress the twofold oscillation in $H_{\rm c2}$, 
it is necessary to satisfy the relations 
\begin{eqnarray} && 
 \langle (\hat{v}_a^2 -\hat{v}_b^2)
  \overline{\phi_{1}^\ast \phi_m}  \rangle_{\mib{k}}
=\langle \hat{v}_a \hat{v}_b 
   \overline{\phi_{1}^\ast \phi_m}  \rangle_{\mib{k}}
\nonumber \\ && 
=\langle \hat{v}_a \hat{v}_c 
   \overline{\phi_{1}^\ast \phi_m}  \rangle_{\mib{k}}
=\langle \hat{v}_b \hat{v}_c 
   \overline{\phi_{1}^\ast \phi_m}  \rangle_{\mib{k}}
=0
\label{eq:zero}
\end{eqnarray} 
in eq. (\ref{eq:hc2e2}), indicating that all gradient coupling terms 
in the GL theory vanish. 
If we consider the correction terms with a higher order of $\gamma$, 
other terms with higher order combination of ${\hat{\mib{v}}}$ appear  
as constraints to be zero. 
The relations in eq. (\ref{eq:zero}) give strong restriction for 
identifying the pairing function. 
For singlet pairing, it is difficult to satisfy all relations in 
eq. (\ref{eq:zero}), when $\phi_1$ is fourfold-symmetric and 
$\phi_m$ is twofold-symmetric. 
For the triplet pairing case, if $\mib{{d}}_1 \cdot \mib{{d}}_m^\ast \ne 0$, 
the difficulty in satisfying eq. (\ref{eq:zero}) is the same as  in the 
singlet pairing case. 
However, all relations in eq. (\ref{eq:zero}) are fulfilled 
in the case of $\mib{{d}}_1 \cdot \mib{{d}}_m^\ast =0$, i.e., 
$\mib{{d}}$-vectors of the triplet pairing functions are 
orthogonal to each other. 

We now discuss the pairing function appropriate 
for ${\rm PrOs_4Sb_{12}}$. 
The possible components of the pairing functions are 
classified under the crystal group consideration.
For the present cubic symmetry ($O_h$), complete classification was 
carried out by Volovik and Gor'kov~\cite{Volovik} in the strong spin-orbit 
coupling case, while Ozaki {\it et al.} enumerated possible $p$-wave 
pairing~\cite{Ozaki1} and higher angular momentum states 
($f$-wave, $\cdots$)~\cite{Ozaki2} in the weak spin-orbit coupling case. 
These constitute a list of the possible states allowed under 
$O_h$ for both singlet and triplet cases. 
The present task is to select the possible states from the list of 
classifications under cubic symmetry
in the light of various theoretical and experimental constraints. 
The constraints we require on the H-phase are 
(i) no gradient coupling by satisfying eq. (\ref{eq:zero}), 
(ii) point nodes on the $a$, $b$ and $c$ axes,\cite{Izawa,Maki}  and 
(iii) $O_h$-symmetric pairing function.

Let us start with the singlet category. 
There are several states with point nodes. 
For example, 
$\phi_1(\mib{{k}})=k_a^2 + \varepsilon k_b^2 +\varepsilon^2 k_c^2$ 
$(\varepsilon={\rm e}^{\pm{\rm i}2 \pi/3})$ has point nodes, but 
along $(1,1,1)$ and equivalent directions. 
It breaks constraint (ii). 
If we consider ad hoc states accidentally combined with higher angular 
momentum states, we can find other point node functions  
satisfying (ii),  
such as the $s+g$-wave state given in eq. (\ref{eq:sg1}). 
However, for these singlet states, the relations in eq. (\ref{eq:zero}) are 
not satisfied when we consider the twofold-symmetric second component. 
Therefore, generally, there is no appropriate state among the 
singlet category free from constraint (i). 

As for the strong spin-orbit triplet category classified by Volovik 
and Gor'kov,\cite{Volovik,Sigrist} the pairing function with point nodes is   
$\mib{{d}}_1(\mib{{k}})
=\hat{x}d_{1,x}+\hat{y}d_{1,y}+\hat{z}d_{1,z}
=\hat{x}k_a+\hat{y}\varepsilon k_b +\hat{z} \varepsilon^2 k_c$   
in the $p$-wave pairing. 
This is a non-unitary state, where one branch of the energy gap 
$|\mib{{d}}_1|^2 \pm |\mib{{d}}_1 \times\mib{{d}}_1^\ast|$ is a full gap, 
and the other has 8 point nodes along $(1,1,1)$ and  
equivalent directions, contrary to constraint (ii). 
In $f$-wave pairing, we have 
$\mib{{d}}_1(\mib{{k}})=\hat{x}k_a (k_c^2 -k_b^2)
+ \hat{y} k_b(k_a^2 -k_c^2) + \hat{z} k_c(k_b^2 -k_a^2)$ and 
$\mib{{d}}_1(\mib{{k}})=\hat{x}k_a (k_c^2 -k_b^2)
+ \hat{y} \varepsilon k_b(k_a^2 -k_c^2) 
+ \hat{z} \varepsilon^2 k_c(k_b^2 -k_a^2)$. 
These have 14 point nodes, i.e., 
8 point nodes along $(\pm 1,\pm 1,\pm1)$ in addition to 
6 point nodes on the $a$, $b$ and $c$ axes, 
which do not agree with the experimental data when the magnetic  
field is tilted from the $ab$-plane.\cite{Izawa,Maki} 
Furthermore, since $\mib{{d}}_1$ has all $d_x$, $d_y$ and $d_z$ components 
in the strong spin-orbit coupling case for $O_h$ symmetry, 
generally $\mib{{d}}_1 \cdot \mib{{d}}_m\ne 0$ 
for the second component $\mib{{d}}_m$,  
and it is difficult to satisfy constraint (i).  
Only when $\mib{{d}}_m \propto \mib{{C}}(\mib{{k}})\times\mib{{d}}_1^\ast $ 
with arbitrary vector function  $\mib{{C}}(\mib{{k}})$, 
we can satisfy  $\mib{{d}}_1 \cdot \mib{{d}}_m^\ast  = 0$.  
However, even when the second component $\mib{{d}}_m$ appears, 
all point nodes for  $\mib{{d}}_1$ do not vanish in the $f$-wave cases, 
which is contrary to that suggested by the experimental data.\cite{Izawa,Maki} 
Thus, there is no suitable state in this category. 

In the last category of the weak spin-orbit coupling triplet state, 
we can freely select the spin part. 
However, other than the pairing functions already listed in the strong 
spin-orbit coupling case, there are no suitable functions which have 
point nodes in the $O_h$ symmetry among the list of the classifications. 
However, if we consider the ad hoc state, 
we can obtain some states to simultaneously satisfy (i)-(iii) such as 
\begin{eqnarray} 
\mib{{d}}_1(\mib{{k}})=\hat{x}(k_a\pm{\rm i}k_b)
(k_b\pm{\rm i}k_c)(k_c\pm{\rm i}k_a) , 
\label{eq:d2}
\end{eqnarray} 
which has 6 point nodes on the $a$, $b$, $c$ axes. 
It can be generally concluded that we have to take a broken time-reversal 
symmetry state (i.e., complex function) to have point nodes 
in triplet pairing. 

Having found possible states for the H-phase, such as in eq. (\ref{eq:d2}), 
we now discuss the L-phase, which should reduce the symmetry from fourfold 
to twofold in the $ab$-plane. 
Below the second transition, the pairing function $\mib{{d}}_m$ 
satisfying  $\mib{{d}}_1 \cdot \mib{{d}}_m^\ast  =0$ 
starts to grow as a second-order transition, 
which breaks the fourfold-symmetry in the $ab$-plane. 
This can be possible if we choose, for example, 
$\mib{{d}}_m(\mib{{k}})=\hat{y}k_a$ or  
$\mib{{d}}_m(\mib{{k}})=\hat{y}(k_a\pm{\rm i}k_c)$.
The 4 point nodes on the $a$ and $b$ axes in the H-phase changes into 
2 point nodes only on the $b$-axis, reducing to twofold-symmetry 
in the L-phase. 
It is noted that the L-phase is necessarily non-unitary, where time-reversal 
symmetry is broken in both the orbital and spin parts of the pairing function, 
because $\mib{{d}}\times\mib{{d}}^\ast \ne0$ for 
$\mib{{d}}=\mib{{d}}_1+\mib{{d}}_2$, 
as $\mib{{d}}_1$ is a complex function. 
In this case, one branch of the energy gap 
$|\mib{{d}}|^2 \pm |\mib{{d}}\times\mib{{d}}^\ast|$  
of the non-unitary state has a small gap, but it remains finite 
because $d_x \ne d_y$ in this case. 
Therefore, the half-residual density of states in the non-unitary states 
vanishes at low temperature.  
In this L-phase, we expect anisotropic spin susceptibility because 
the spin component is not isotropic. 
In the non-unitary state, we expect the spontaneous moment in the spin part 
of the pairing state, 
which can be a key point to determining the pairing function. 

In summary, 
for the fourfold (H-phase)-twofold (L-phase) transition in 
${\rm PrOs_4Sb_{12}}$, we have examined the mixing effect due to 
gradient coupling between multicomponents in unconventional 
superconductors by calculating $H_{\rm c2}$ near $T_{\rm c}$. 
We then narrow down the possible pairing functions in both the  
H- and L-phases of ${\rm PrOs_4Sb_{12}}$ in light of the existing 
experimental information. 
If we require that the constraints (i)-(iii) mentioned above must be 
fulfilled simultaneously, the candidates for the H-phase are the 
pairing states, such as those in eq. (\ref{eq:d2}). 
The main points of our results are:  
(A) 
Any spin-singlet states are excluded due to gradient coupling, 
if the second transition is of the second order. 
The weak spin-orbit coupling triplet pairing state is plausible 
as a pairing function. 
(B) 
In the H-phase, 
it is most likely a broken time-reversal symmetry state in the orbital part,  
because of point nodes in triplet pairing. 
In addition, the L-phase becomes necessarily a non-unitary state 
by the combination of two order parameters. 

The authors would like to thank Y. Aoki, K. Izawa and 
Y. Matsuda for helpful discussion.


\end{document}